\documentclass[twocolumn]{aastex7}

\usepackage{subcaption}
\begin{document}

\title{Evidence of Long-Lived Powerful Gyrosynchrotron Radio Emission in the Close Binary FF~UMa}
\author[0009-0003-1870-5164]{Ruijie Gao}
\affiliation{School of Physics and Astronomy, Yunnan University, Kunming 650091, P. R. China}
\email[show]{ruijiegz@gmail.com}  

\author[0000-0002-2322-5232]{Jun Yang}
\affiliation{Department of Space, Earth and Environment, \\Chalmers University of Technology, Onsala Space Observatory, 43992 Onsala, Sweden}
\email{jun.yang@chalmers.se}

\author[0000-0002-6316-1632]{Yang Gao}
\affiliation{School of Physics and Astronomy, Sun Yat-Sen University, Zhuhai, 519082 Guangdong, P. R. China} 
\email{gaoyang25@mail.sysu.edu.cn}

\author[0000-0002-9768-2700]{Jingdong Zhang}
\affiliation{Shanghai Astronomical Observatory, Chinese Academy of Sciences, 80 Nandan Road, Shanghai, P. R. China} 
\affiliation{Department of Geodesy and Geodynamics Finnish Geospatial Research Institute (FGI), National Land Survey of Finland, Vuorimiehentie 5, Espoo 02150, Finland}
\email{zhangjingdong@shao.ac.cn}

\author[0000-0003-1353-9040]{Bo Zhang}
\affiliation{Shanghai Astronomical Observatory, Chinese Academy of Sciences, 80 Nandan Road, Shanghai, P. R. China} 
\email{zb@shao.ac.cn}

\author[0000-0002-5519-0628]{Wen Chen}
\affiliation{Yunnan Observatories, Chinese Academy of Sciences, Kunming 650216, Yunnan, P. R. China} 
\email{chenwen@ynao.ac.cn}

\author[0000-0002-3464-5128]{Xiaohui Sun}
\affiliation{School of Physics and Astronomy, Yunnan University, Kunming 650091, P. R. China}
\email{xhsun@ynu.edu.cn}

\author[0000-0001-6383-6751]{Guannan Gao}
\affiliation{Yunnan Observatories, Chinese Academy of Sciences, Kunming 650216, Yunnan, P. R. China} 
\affiliation{Yunnan Key Laboratory of the Solar Physics and Space Science, Kunming 650216, P. R. China} 
\email{ggn@ynao.ac.cn}

\author[0000-0002-4280-6630]{Zhibin Dai}
\affiliation{Yunnan Observatories, Chinese Academy of Sciences, Kunming 650216, Yunnan, P. R. China} 
\affiliation{Key Laboratory for the Structure and Evolution of Celestial Objects, Chinese Academy of Sciences, Kunming 650216, P. R. China} 
\email{zhibin_dai@ynao.ac.cn}

\author[0000-0002-4963-179X]{Tobia D. Carozzi}
\affiliation{Department of Space, Earth and Environment, \\Chalmers University of Technology, Onsala Space Observatory, 43992 Onsala, Sweden}
\email{tobia.carozzi@chalmers.se}

\correspondingauthor{Ruijie Gao, Jun Yang, Xiaohui Sun}
 
\begin{abstract}
RS~Canum~Venaticorum~(RS~CVn) close binaries, characterized by tidal locking, rapid rotations, and strong magnetic fields, are ideal laboratories for high-resolution radio observations to probe emission processes, magnetic field configurations, and interaction activity. Despite their importance, only a few RS~CVn sources have been explored by polarimetric observations of very long baseline interferometry~(VLBI). To expand the effort, we have analyzed the existing Very Long Baseline Array~(VLBA) astrometric data for the RS~CVn binary FF~Ursae~Majoris~(FF~UMa). In the 5-GHz VLBA experiments conducted between 2021 and 2024, both total intensity and circularly polarized emission were clearly detected at six of seven epochs. The consistently high brightness temperatures ($\geq10^7$ K) and the moderate fractional circular polarization (10\%--30\%) over about three years indicate that the radio emission is mainly produced by gyrosynchrotron radiation from mildly relativistic electrons in the highly-ordered magnetic field. The radio luminosities are also comparable to those of previously studied powerful RS~CVn binaries and show a significant anti-correlation with fractional circular polarization. A mean centroid offset of $13.4 \pm 3.1~R_\odot$ between the Stokes $I$ and $V$ emission was found across multiple epochs, indicating a possible additional contribution from the secondary star via a magnetically active corona, a giant magnetic loop, or significant interaction activity with the primary star in the quiescent state.
\end{abstract}


\section{Introduction}
RS Canum Venaticorum~(RS~CVn) binary systems are close binaries consisting of a subgiant or giant star of spectral type F, G, or K and a cooler main-sequence companion~\citep[e.g.,][]{Hall1976,Hall1981}. Because of their strong magnetic activity, these systems are important targets for exploring stellar astrophysics. The short orbital period (1--30 days) leads to tidal synchronization and thus rapid rotation of both stellar components. Combined with their deep convective envelopes, this rapid rotation drives strong magnetic dynamos that power intense chromospheric and coronal activity.~\citep{Ayres1980}. Therefore, RS~CVn systems can be detected across a wide range of wavelengths, from radio to high-energy bands. As such, RS~CVn stars are often regarded as natural laboratories for high-resolution imaging observations to study stellar magnetic field structures and activities. 

 RS~CVn systems have been extensively observed in the radio band since the late 1970s~\citep{Spangler1977, Mutel1985, Morris1988, Drake1989}. These active binaries produce radio emission through magnetically driven non‑thermal processes in their active stellar coronae. The dominant mechanism at the GHz frequency range is likely incoherent gyrosynchrotron radiation from mildly relativistic electrons spiraling in magnetic fields of tens to hundreds of gauss~\citep{Dulk1985,G_B2002,2026PASA}. This explains the non-thermal radio spectra and significant variability. To date, most observations have been conducted with single-dish telescopes or low-resolution interferometers, limiting our ability to resolve the spatial structure of the emission regions and the magnetic fields~\citep{2025PASP}. Just a few RS~CVn binaries have been observed using Very Long Baseline Interferometry~(VLBI), with notable examples including UX~Ari, HR~1099, and AR~Lac~\citep[e.g.,][]{Mutel1985, Mutel1987, Massi1988AA,CW_HR1099}. High-resolution VLBI studies have revealed complex morphologies in total intensity and circularly polarized emission across multiple wavelengths, providing crucial insights into the geometry of magnetically active regions.

 The dominant stellar component responsible for radio emission in RS~CVn binaries remains uncertain. Radio emission in these systems is often assumed to be dominated by the primary star, owing to its stronger magnetic activity~\citep{Brun2017} in the corona and the presence of potential giant magnetic loops. HR~1099, being the closest RS~CVn binary to Earth, has become the best-studied system in terms of long-term, multi-epoch VLBI observations~\citep[e.g.,][]{Lestrade1984ApJL, Mutel1984ApJ, Mutel1985ApJ, Massi1988AA}. With the advent of the Very Long Baseline Array~(VLBA), \citet{Ransom2002ApJ} obtained high-resolution 8.4~GHz images, resolving a double-peaked radio structure during a flaring state and providing evidence that the radio-emitting regions corotate with the binary on timescales of hours. More recently, \citet{Golay2024ApJ} conducted multi-epoch VLBA astrometric observations and found that the radio emission centroid of HR~1099 is systematically offset from the binary center of mass at all observed epochs, with its positional behavior indicating that the radio emission is predominantly associated with the primary star.

 In some RS~CVn systems, both stellar components may contribute to the radio emission. For example, \citet{Mutel1985ApJ} reported that UX~Ari exhibits a characteristic core--halo radio morphology, with the compact core associated with the primary star and the extended halo linked to the secondary star. In later VLBI observations of UX~Ari, however, the giant halo was not detected \citep[e.g.][]{Peterson2011ApJ}. 

 A different case suggests that the secondary star may dominate the radio emission. Using multi-epoch VLBA observations at 15~GHz, \citet{Abbuhl2015ApJ} carried out phase-referenced radio astrometry of the RS~CVn binary system HR~5110. After accounting for proper motion, parallax, and orbital motion, they found that the radio emission centroid is consistently coincident with the K-type secondary star. This result supports an interpretation in which the radio emission in HR~5110 is dominated by the KIV secondary component rather than the FIV primary star or an interbinary interaction region. To date, long-term VLBI detections of both stellar components have not been reported in RS~CVn binaries.

 Among known RS~CVn systems, FF Ursae Majoris~(FF~UMa, also known as 2RE~J0933+624 or HD~82286) stands out as a particularly active binary that has received some prior study. Because of significant chromospheric activity discovered in the ROSAT EUV all-sky survey by \citet{Pounds1993,Vedantham2022}, it was classified as a close binary with a period of $\sim$3.27~d~\citep{Jeffries1995, Griffin2012}. The accurate orbital parameters of FF~UMa are summarized in Table~\ref{tab:orbital parameters}. As a rapidly rotating and tidally locked binary, both stellar components show rarely-seen strong H$\alpha$ emission and photometric variability, indicating extraordinary magnetic activity~\citep{Strassmeier2000, Galvez2007, Senavci2020}. FF~UMa has also been detected in several wide-field radio surveys, including the LOFAR Two-metre Sky Survey \citep[LoTSS,][]{LoTSS}, the Faint Images of the Radio Sky at Twenty Centimeters survey \citep[FIRST,][]{FIRST}, the NRAO VLA Sky Survey \citep[NVSS,][]{NVSS}, and the VLA Sky Survey \citep[VLASS,][]{VLASS}. However, these radio observations were obtained with limited angular resolution and at sparse epochs, leaving the spatial origin and geometry of the radio emission largely unconstrained. High-resolution, multi-frequency VLBI observations are therefore essential to investigate the spatial distribution of the radio emission, to identify the dominant radiation mechanisms at GHz frequencies, and to constrain the geometry and strength of the magnetic fields in FF~UMa.

 \begin{deluxetable}{lcc}
\digitalasset 
\tablewidth{\columnwidth}
\tablecaption{\textbf{Orbital parameters of FF~UMa} \label{tab:orbital parameters}}
\tablehead{\colhead{Parameter}  & \colhead{Symbol and Value} &\colhead{References}}
\startdata
Spectral type        &  K1IV + K0V & (1) \\
Primary radius       & $r_\mathrm{P} = 2.95\mathrm{R}_{\odot}$ & (2) \\
Secondary radius     & $r_\mathrm{S} = 2.30\mathrm{R}_{\odot}$ & (2) \\
Parallax             & $\pi = 9.57$mas  & (3)\\
Eccentricity         & $e = 0.0$ & (1) \\
Mass ratio           & $q = 2.12\pm0.10$ & (1)\\
Orbital inclination  & $i_\mathrm{obs} = 50.5^\circ$ & (2) \\
Orbital period   & $P = 3.27487\pm0.00004$~d  & (2) \\
Ephemeris epoch (MJD) & $T_\mathrm{eph}=54067.019\pm0.0017$ &(2) \\
\enddata
\tablecomments{Reference.(1)~\cite{Galvez2007}; (2)~\cite{Senavci2020}; (3)~\cite{table1_3}.}
\end{deluxetable}

In this study, we reanalyzed the existing VLBA data of FF~UMa. By combining these data with the well-constrained orbital and rotational parameters of the system, we determined the degree and handedness of polarization, assessed the underlying emission mechanisms, and investigated the spatial structure of the binary’s magnetic field. Section~\ref{sec:observations} describes the observations and data processing; Section~\ref{sec:results} presents our results; Section~\ref{sec:discussion} discusses these findings; and Section~\ref{sec:conclusion} concludes the paper.

\section{VLBA observations and data reduction}
\label{sec:observations}
FF~UMa was observed with the VLBA in 2021 and 2024 under the VLBA project codes BZ087~(PI: Bo Zhang), BZ103~(PI: Jingdong Zhang), and BZ107~(PI: Jingdong Zhang) at 4.8~GHz. These projects were mainly designed to validate the Gaia Celestial Reference Frame via VLBI astrometry \citep{CW2023MN,ZJD2024MN,ZJD2025AA,ZJD2026MN}. 

Because these experiments were performed in the phase-referencing mode (cycle observations of FF~UMa and nearby calibrators), their data could also allow us to search for highly circular polarized emission from the weak radio star FF~UMa without the standard calibration of small (about a few percent) instrumental cross-polarization leakages \citep[e.g.][]{Homan1999AJ}. In total, seven epochs were observed; the date and participating antennas for each epoch are listed in Table~\ref{tab:observation_parameters}. Each epoch included 9--10 VLBA antennas, recording at 4096~Mbps (4 intermediate frequencies (IFs) in dual circular polarization, each 128~MHz wide, with 2-bit quantization). FF~UMa was observed for $\sim$40--45~minutes per epoch, with a phase-referencing cycle time of 4~minutes. The raw data were correlated with the VLBA DiFX software correlator \citep{2007PASP} using 2~s integration and 0.5~MHz frequency resolution.

The original observations were designed following the MultiView strategy \citep{Rioja2017AJ,ZJD2026MN}. Given the small angular separation between FF~UMa and the calibrator J0921$+$6215 (from the ICRF3 catalog), the phase solutions derived from this source sufficiently capture atmospheric effects, making the data reduction effectively equivalent to standard phase referencing. \citet{ZJD2026MN} compared calibration using this single phase calibrator with the MultiView approach and found that the resulting astrometric uncertainties are at a comparable level. During the observations, this phase calibrator had an average flux density of $\sim0.46$~Jy.

Visibility data were calibrated and imaged using the National Radio Astronomy Observatory (NRAO) software package Astronomical Image Processing System~\cite[AIPS;][]{Greisen2003}. The calibration steps followed the sequence: First, we did minor amplitude corrections in cross-correlation data due to errors in sampler thresholds using measurements of auto-correlation data. Second, system temperature data and antenna gain curves were used to perform a priori amplitude calibration. Third, the phase errors resulting from the changes in the parallactic angle were corrected. Fourth, the Earth Orientation Parameters (EOP) were updated according to the EOP products provided by the United States Naval Observatory (USNO), and the related phase errors were corrected. Fifth, the dispersive delays due to the propagation effect in the Earth ionosphere were derived and removed with the Global total electron content maps provided by the Global Navigation Satellite System (GNSS) community. Sixth, the AIPS task \texttt{FRING} was first applied to the bright calibrator B0217$+$734 for manual phase calibration. After applying these solutions, we combined all IFs and both Stokes $RR$ and $LL$ to perform a global fringe fitting on the phase-referencing calibrator J0921$+$6215. To gain accurate phase interpolations from calibrator to target scans, we smoothed fringe rate solutions. Finally, we applied the accumulated complex gain solutions and split the data into single-source files. We did deconvolution and self-calibration on the phase-referencing calibrator with the AIPS tasks \texttt{IMAGR} and \texttt{CALIB}. The calibrator shows a core–jet structure with a bright core and a faint, tail-like, short jet toward the northwest. According to the non-thermal synchrotron emission mechanism and statistical research on radio sources \citep[e.g.][]{Homan2006AJ}, its circular polarization fraction is expected to be very low ($<1$\%) and can be safely used as a zero-$V$ calibrator for self-calibration. Using the final source model, we also re-run self-calibration with the multi-source data file and then transferred both amplitude and phase solutions to the target source. After applying the final solutions, we imaged FF~UMa with the default Briggs robust weighting parameter of zero and without any self-calibration.
 
\begin{deluxetable*}{ccccccccc}
\digitalasset
\tablewidth{0pt}
\caption{\textbf{Dates and participating antennas of the VLBA observations}}
\label{tab:observation_parameters}
\tablehead{
\colhead{Epoch} &
\colhead{Project Code} &
\colhead{Date} &
\colhead{MJD} &
\colhead{Phase} &
\colhead{Absent Stations} &
\colhead{Beam} &
\colhead{Beam} &
\colhead{Beam}  \\
 & & (yyyy-mm-dd) & (mid-epoch) & & & Maj(mas) & Min(mas) & PA($^\circ$) \\
 \colhead{(1)} & \colhead{(2)} & \colhead{(3)} & \colhead{(4)} & \colhead{(5)} & \colhead{(6)} & \colhead{(7)} & \colhead{(8)} & \colhead{(9)}
}
\startdata
1 & BZ087D1 & 2021-10-09 & 59496.6745 & 0.128 & PT & 3.65 & 1.22 & 6.86 \\
2 & BZ087D2 & 2021-12-06 & 59554.5161 & 0.791 & MK & 3.25 & 1.68 & -7.03 \\
3 & BZ103D1 & 2024-04-14 & 60414.1626 & 0.289 & -- & 2.69 & 1.17 & -7.93 \\
4 & BZ103D2 & 2024-05-06 & 60436.0896 & 0.984 & FD & 2.88 & 1.14 & -0.35 \\
5 & BZ103D3 & 2024-06-05 & 60466.0206 & 0.124 & -- & 2.76 & 1.18 & -7.65 \\
6 & BZ107D1 & 2024-10-18 & 60601.6493 & 0.539 & -- & 2.81 & 1.23 & -11.72 \\
7 & BZ107D2 & 2024-12-07 & 60651.5125 & 0.765 & -- & 2.84 & 1.24 & -11.23 \\
\enddata

\tablecomments{
Column (6) lists the VLBA stations that did not participate in each observing epoch. The VLBA stations are: Brewster (BR), Fort Davis (FD), Hancock (HN), Kitt Peak (KP), Los Alamos (LA), Mauna Kea (MK), North Liberty (NL), Owens Valley (OV), Pie Town (PT), and Saint Croix (SC). Columns (7)–(9) give the major axis, minor axis, and position angle of the CLEAN restoring beam used in the final VLBA images, which are also shown in the lower left corner of each panel in Figure~\ref{fig:IV_images_of_ffuma}.}
\end{deluxetable*}

\section{Detections of Stokes \textit{I} and \textit{V} emission in magnetically active binary FF~Uma} 
\label{sec:results}
\begin{figure*}[ht!]
\centering  
\newcommand{\figw}{0.25\linewidth}
\setlength{\tabcolsep}{2pt} 
\renewcommand{\arraystretch}{0} 

\begin{tabular}{@{}cccc@{}}
\includegraphics[height=0.35\textheight,trim=130 90 65 100,clip]{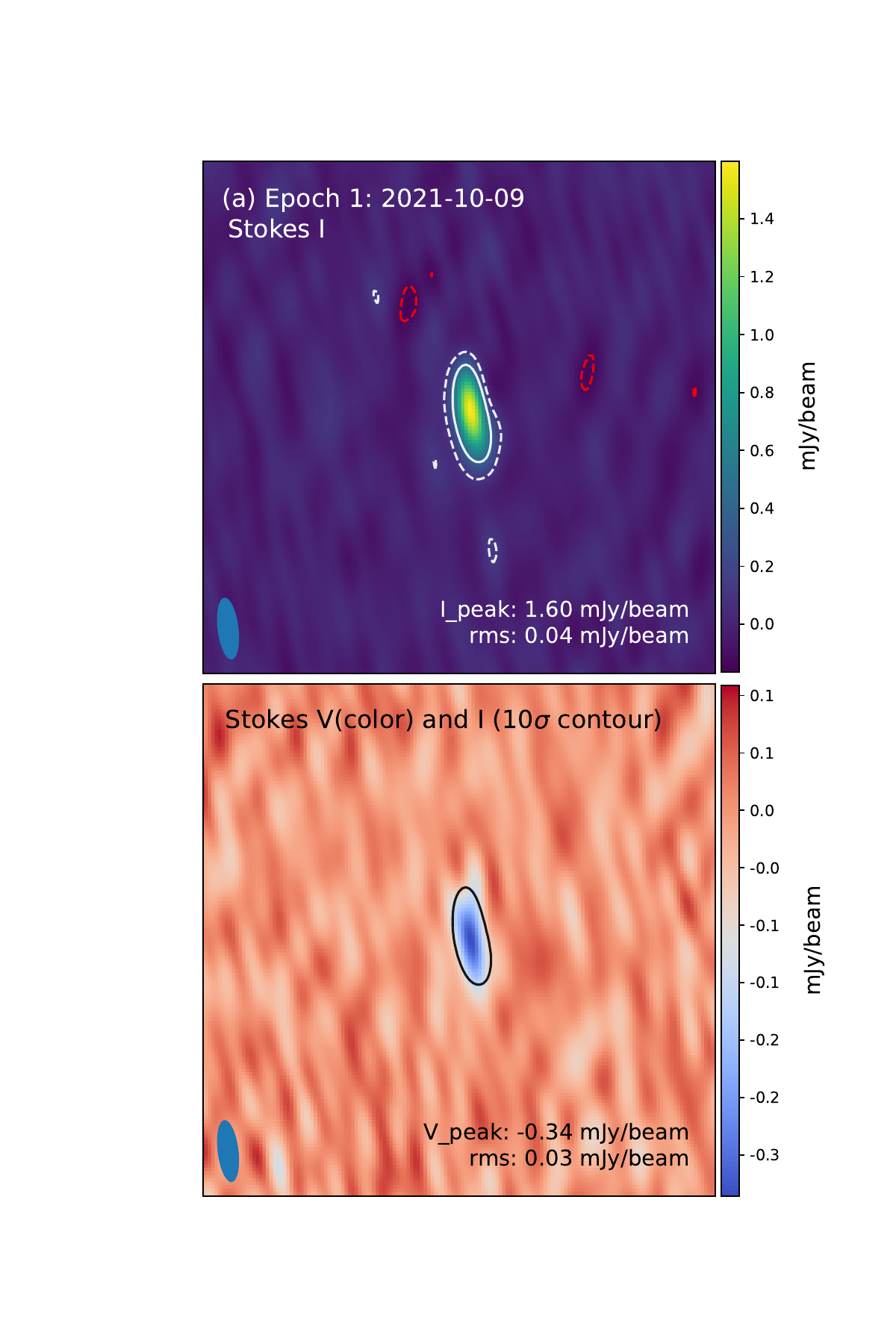} &
\includegraphics[height=0.35\textheight,trim=130 90 65 100,clip]{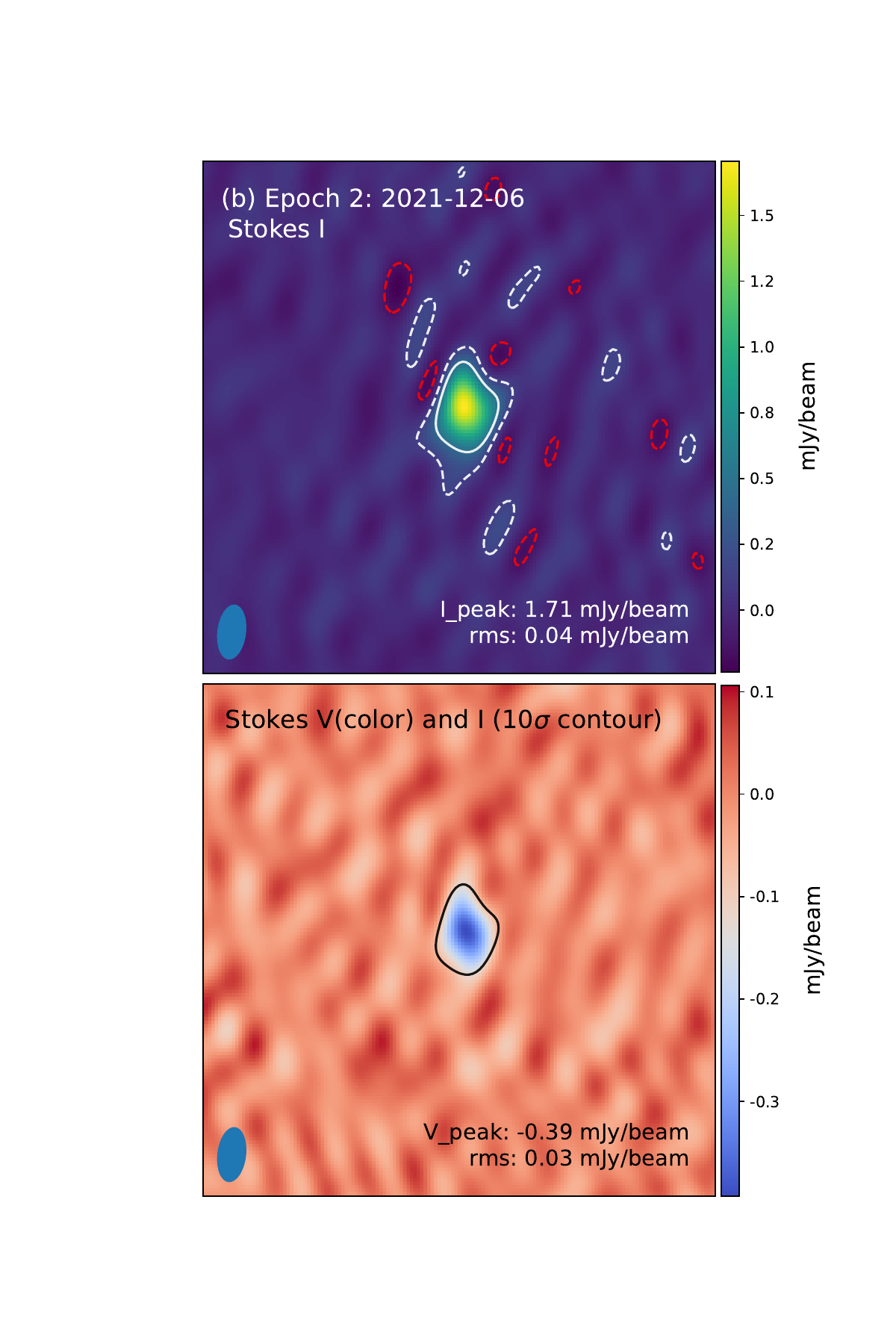} &
\includegraphics[height=0.35\textheight,trim=130 90 65 100,clip]{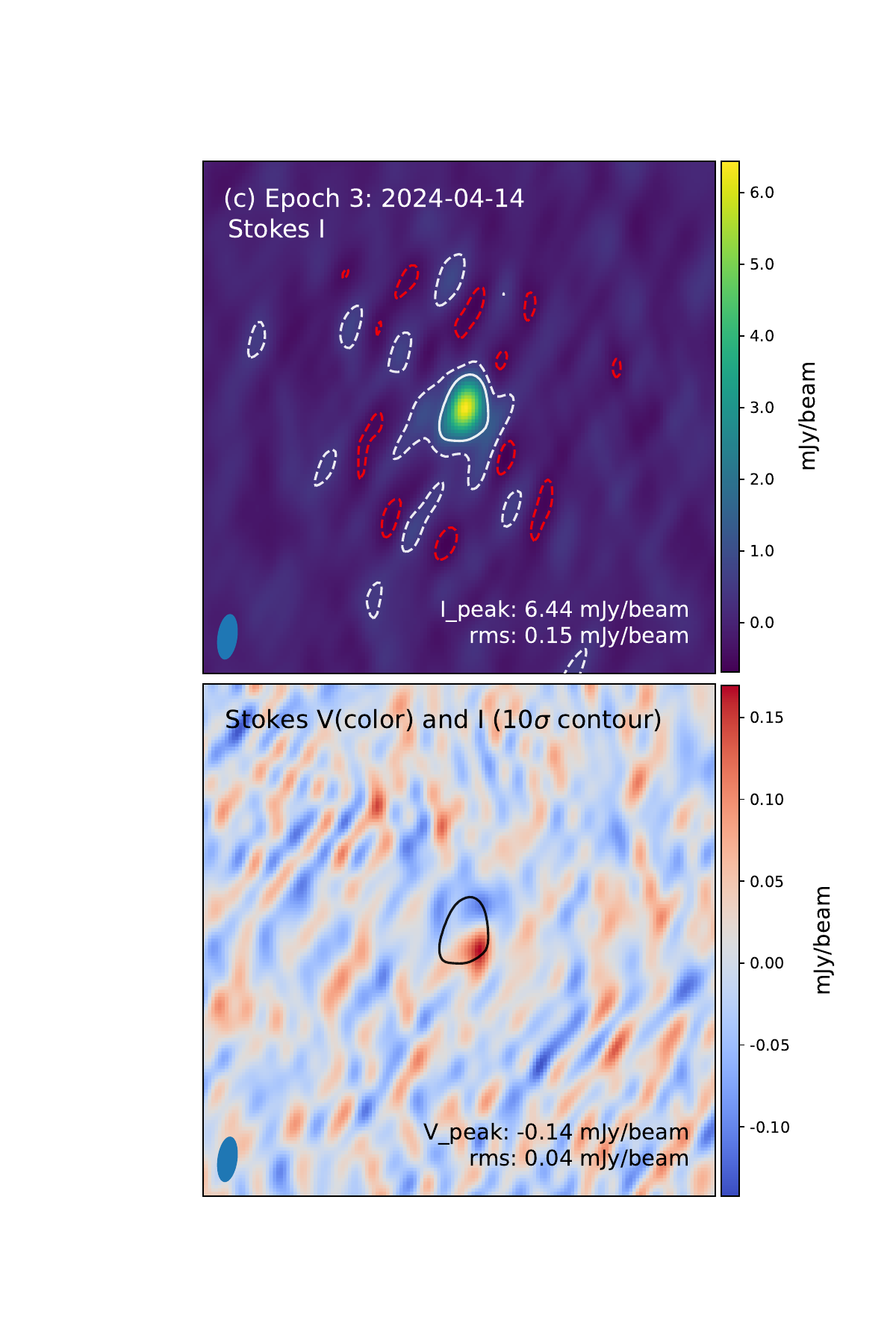} &
\includegraphics[height=0.35\textheight,trim=130 90 45 100,clip]{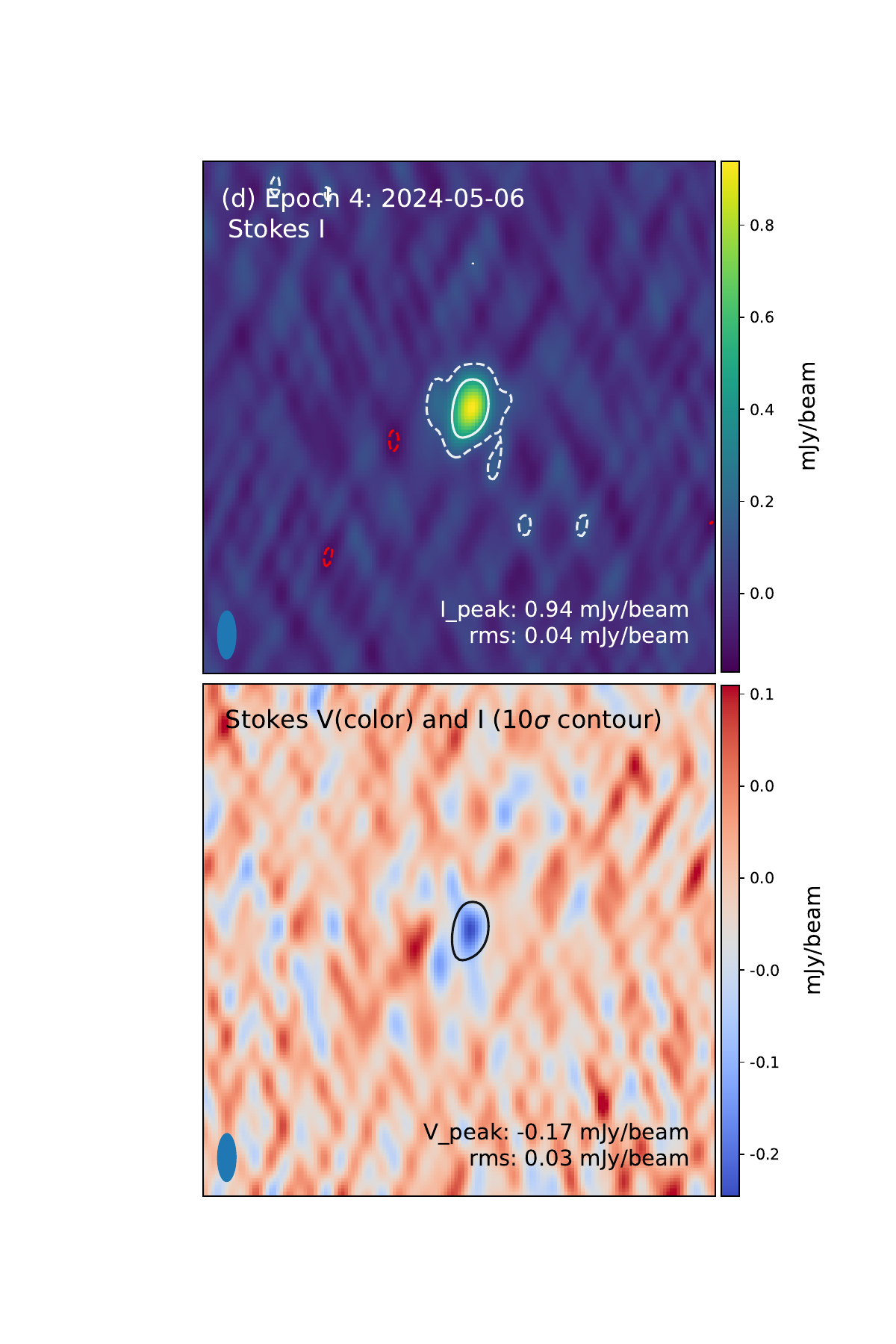} \\
[0.4em]
\includegraphics[height=0.35\textheight,trim=130 90 65 100,clip]{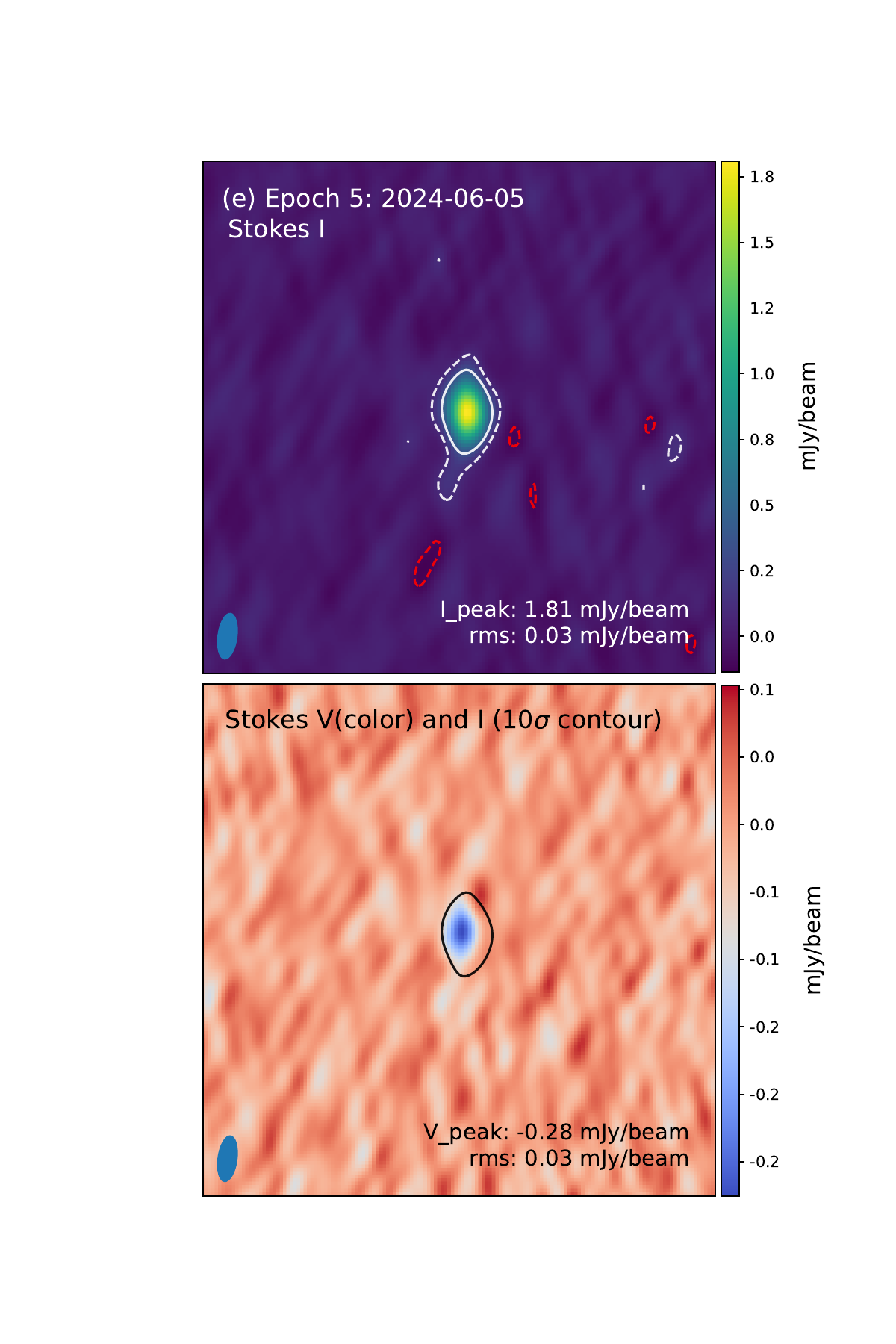} &
\includegraphics[height=0.35\textheight,trim=130 90 65 100,clip]{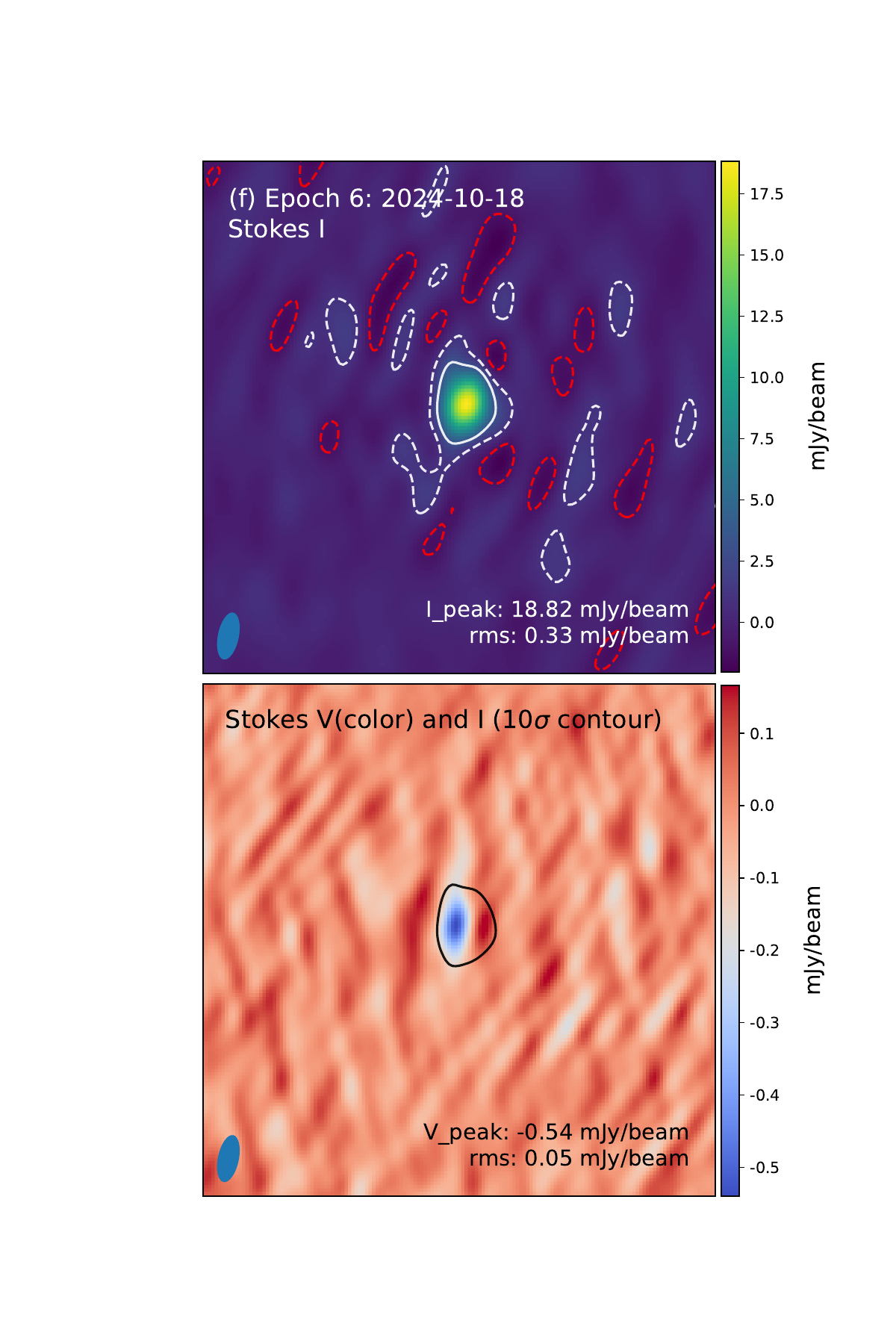} &
\includegraphics[height=0.35\textheight,trim=130 90 45 100,clip]{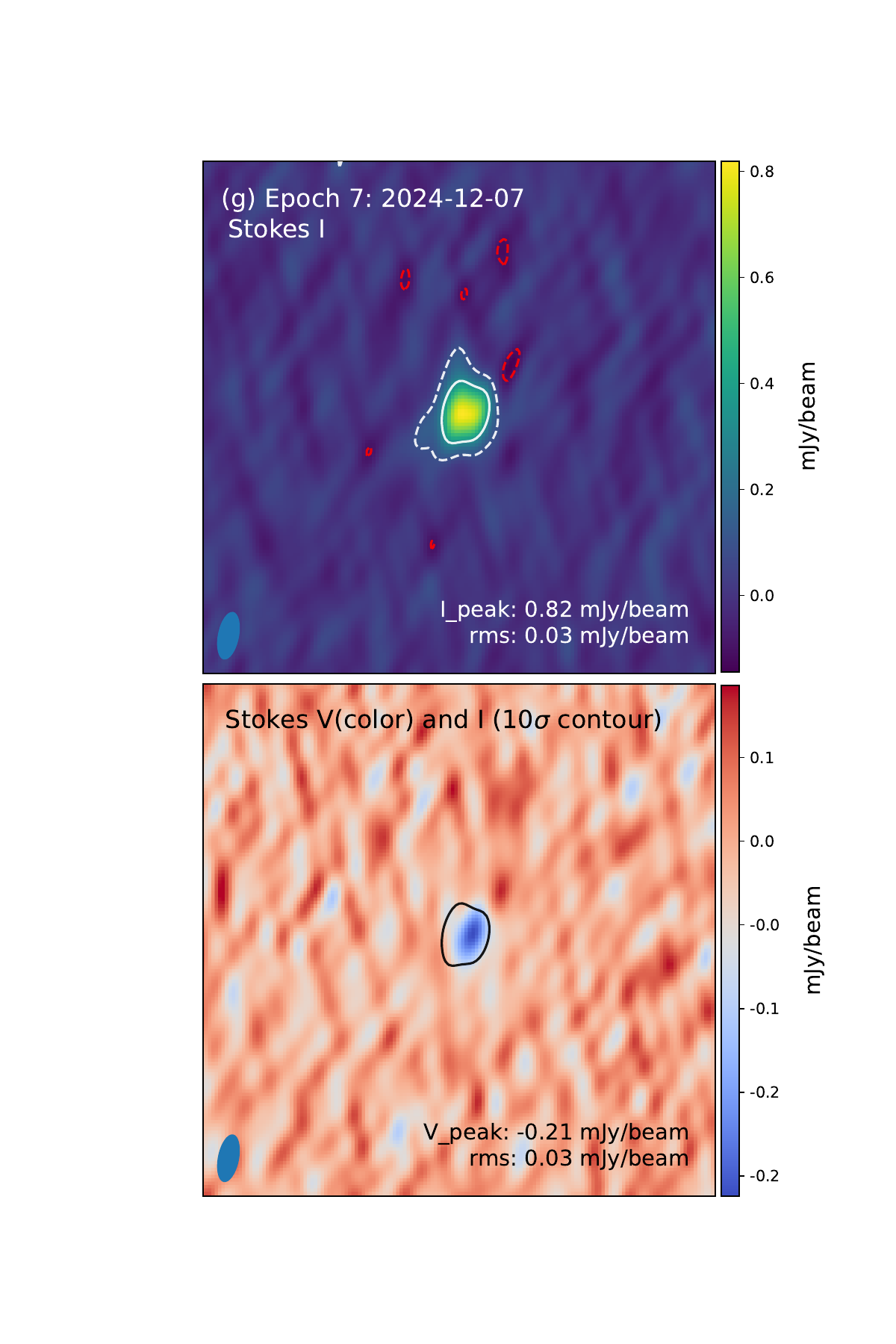} &
{} 
\end{tabular}
\caption{Images of the target source FF~UMa in all epochs. Each panel covers a field of view of $30 \times 30~\mathrm{mas}^2$. The color bar on the right side of each panel is in units of mJy/beam$^{-1}$. In each panel, the upper subpanel shows the Stokes~$I$ image and its $+10\sigma$ (white solid), $+3\sigma$ (white dashed), and $-3\sigma$ (red dashed) contours. The lower subpanel shows the Stokes~$V$ image, overlaid with the Stokes~$I$ $+10\sigma$ (black solid) contour. The synthesized beam is shown in the lower left corner.}
\label{fig:IV_images_of_ffuma}
\end{figure*}

\begin{figure}[htbp]
  \centering
  \includegraphics[height=0.52\textheight,trim=10 0 540 0, clip]{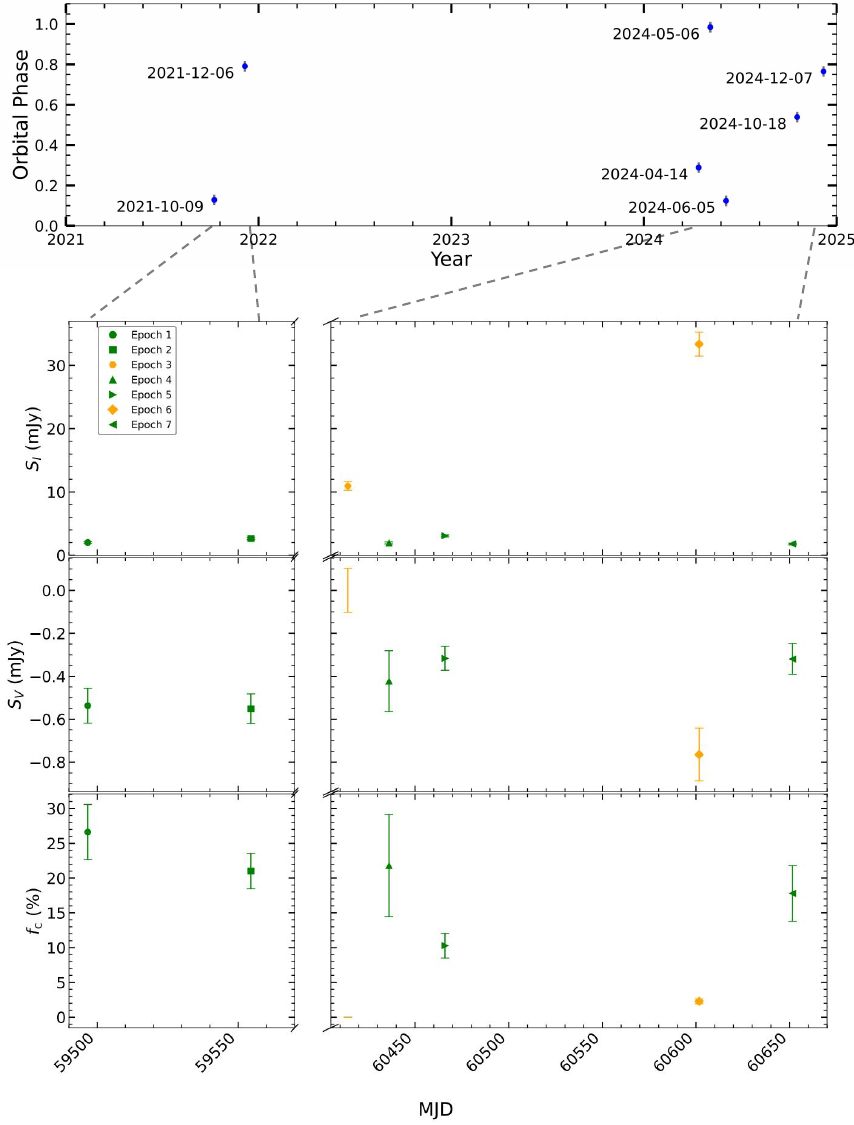}  
  \caption{Top: orbital phase coverage of the multi-epoch VLBA observations of FF~UMa. The observing date and epoch number are labeled next to each point. Bottom: total intensity $S_\mathrm{I}$ (top), circularly polarized flux density $S_\mathrm{V}$ (middle), and fractional circular polarization $f_\mathrm{c}$ (bottom) vs. observing time (MJD). Different marker shapes indicate different observing epochs. Green symbols denote quiescent states, while orange symbols mark flaring epochs.}
  \label{fig:flux_epoch}
\end{figure}

\begin{figure*}[htbp]
    \centering
    \begin{subfigure}[t]{0.5\textwidth}
        \centering
        \includegraphics[width=\linewidth]{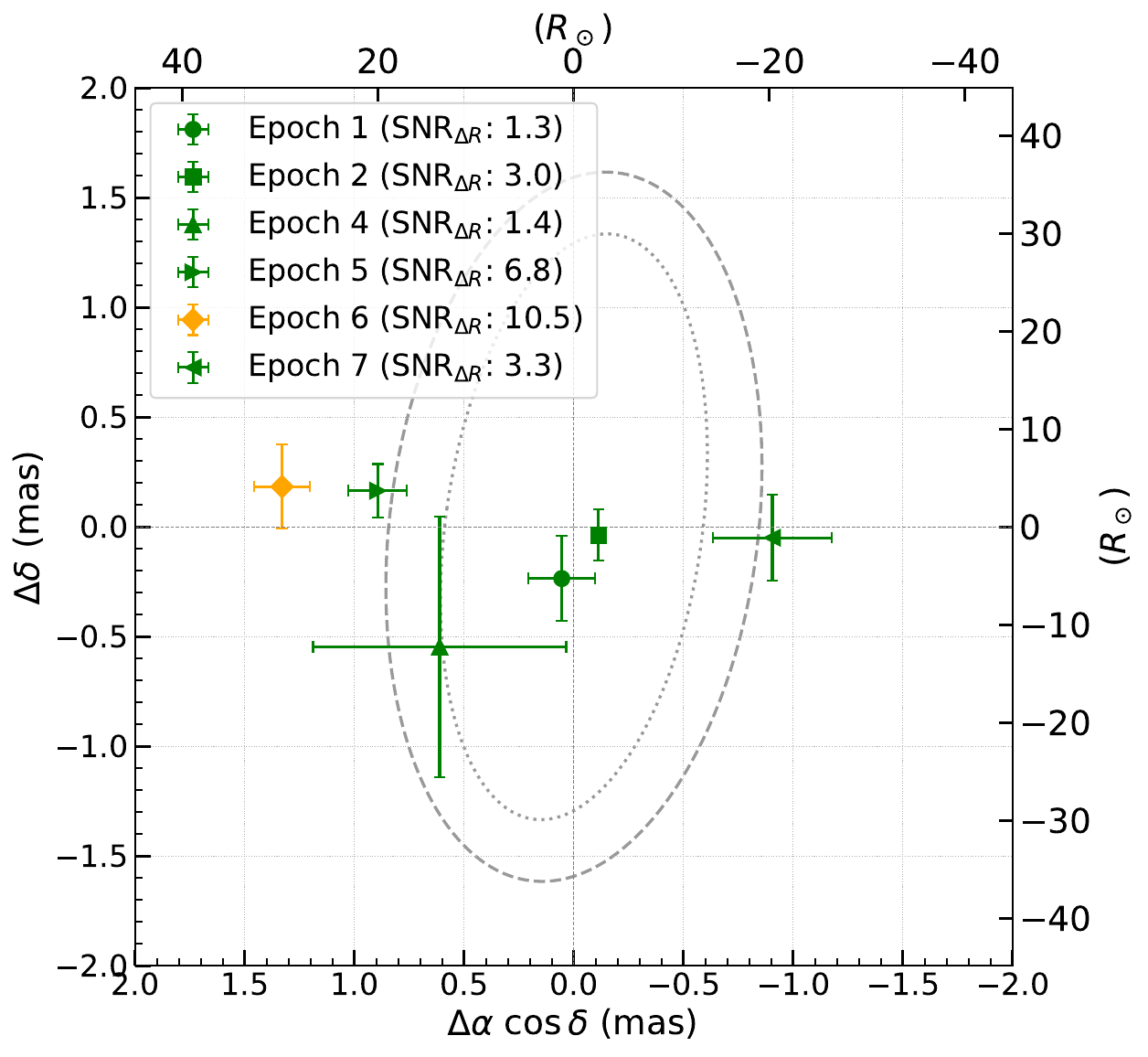}
    \end{subfigure}\hfill
    \begin{subfigure}[t]{0.45\textwidth}
        \centering
        \includegraphics[width=\linewidth]{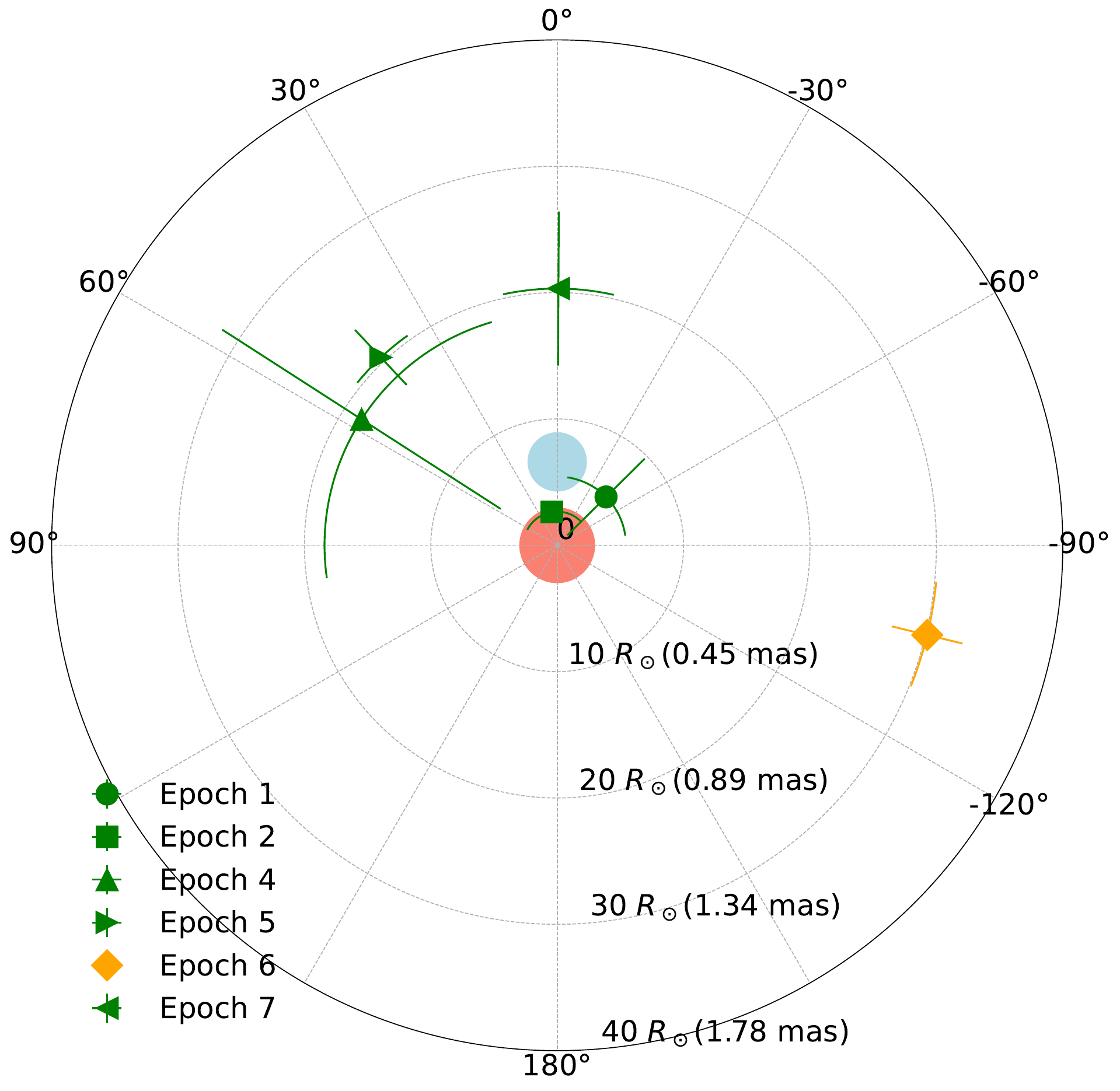}
    \end{subfigure}
    \caption{Positional offsets between the Stokes $I$ and $V$ centroids of FF~UMa. Different symbols denoted different observing epochs. Epoch 6, which corresponded to a flaring state, was shown in orange, while the remaining epochs in quiescent states were shown in green. In both panels, the Stokes $I$ position was taken as the coordinate origin and was not explicitly plotted. Left: Offsets plotted in the sky plane. The SNR of the positional offset for each epoch was indicated in the legend. The gray dashed ellipses denote the synthesized beams with the minimum~(Epoch~3) and maximum~(Epoch~2) areas among all epochs. Right: Offsets plotted in the binary reference frame. Red and blue circles indicated the positions of the primary and secondary stars, respectively; their sizes did not represent the actual physical scales of the stars, $0^\circ$ corresponded to the direction from the primary star toward the secondary star.}
    \label{fig:offsets_result}
\end{figure*}

The target RS~CVn binary FF~UMa was clearly detected with signal-to-noise ratio~(SNR) of 14--18 in all epochs, as shown in Figure~\ref{fig:IV_images_of_ffuma}. Table~\ref{tab:jmfit_result} summarizes the imaging results of Stokes \textit{I} and \textit{V}. Their total flux densities ($S_\mathrm{I}$, $S_\mathrm{V}$), ratios ($f_\mathrm{c}=\frac{|S_\mathrm{V}|}{S_\mathrm{I}}$), and positional offsets are also shown in Figure~\ref{fig:flux_epoch} and Figure~\ref{fig:offsets_result}. These results were derived through fitting an elliptical Gaussian model to the Stokes~$I$ and $V$ maps using the AIPS task \texttt{JMFIT}. The fitted parameters include the source peak brightness, integrated flux density, position, major and minor axis sizes, and the position angle of the major axis. The VLBA observations covered a broad range of orbital phases and had a relatively uniform distribution (c.f. Figure~\ref{fig:flux_epoch}). The maximum gap is approximately 0.2 phase between adjacent epochs.

FF~UMa displays significant variability between epochs in Figure~\ref{fig:flux_epoch}. It might be in the flare state with significantly high flux densities~(10.94$\pm$0.67 and 33.37$\pm$1.90 mJy) in epochs 3 and 6, and in the quiescent state with relatively low and stable flux densities~(1.94--3.07~mJy). Significant circular polarization emission was also detected in 6 of 7 epochs. All detected Stokes~$V$ components show left-handed helicity and relatively stable total flux densities, 0.32--0.76~mJy. Due to non-detection of Stokes \textit{V} emission in epoch 3, only $\pm$3$\sigma$ limits were given. To verify the reliability of the calibration, we also examined the imaging results of the compact calibrator J0921$+$6215. Its total flux density remained relatively stable, varying by less than 10~per~cent across all epochs. No signal exceeding $3\sigma$ in the on-source region was detected in the Stokes~$V$ maps (e.g. Figure~\ref{fig:images_of_J0921} in Appendix~\ref{app:all_IV_fig}). For all seven epochs, these values are very small ($\leq$1.2\%), supporting the conclusion that the flux density variations and large fractional circular polarization (10\%--30\%) observed in FF~UMa are intrinsic to the target source. Uncorrected cross-polarization leakage can convert a fraction of the linear polarization into a spurious circular polarization signal in the data. According to the early investigation by \citet{Homan1999AJ}, this spurious signal is small ($<0.25\%$), even if we assume an unusually high fractional linear polarization of $10\%$ \citep[e.g.,][]{Mutel1984ApJ, Phillips1996AJ, SalterAA2010}. Its impact on the position measurements is therefore not significant ($<1\sigma$) in the Stokes~$V$ maps.

\begin{deluxetable*}{c rr rr rrr r}
\digitalasset
\tablewidth{0pt}
\tablecaption{Summary of VLBA Stokes $I$ and $V$ imaging results of FF~UMa  \label{tab:jmfit_result}}
\tablehead{
\colhead{Epoch-Stokes} &\colhead{Peak}&\colhead{Integrated} & \colhead{RA} & \colhead{Dec} & \colhead{$\theta_{\mathrm{Maj}}$} & \colhead{$\theta_{\mathrm{Min}}$} & \colhead{$\theta_{\mathrm{PA}}$} & \colhead{$T_\mathrm{b}$} \\
  & \multicolumn{1}{c}{($\mathrm{mJy}\,\mathrm{beam}^{-1}$)}
  & \multicolumn{1}{c}{(mJy)}
  & \multicolumn{1}{c}{(mas)}
  & \multicolumn{1}{c}{(mas)}
  & \multicolumn{1}{c}{(mas)}
  & \multicolumn{1}{c}{(mas)}
  & \multicolumn{1}{c}{($^\circ$)}
  & \multicolumn{1}{c}{($10^7\mathrm{K}$)} \\
\colhead{(1)} & \colhead{(2)} & \colhead{(3)} & \colhead{(4)} & \colhead{(5)} & \colhead{(6)} & \colhead{(7)} & \colhead{(8)} & \colhead{(9)} }
\startdata
1-I & $1.59\pm 0.09$ & $2.02\pm0.13$  & $0.0000\pm0.03$ & $0.0000\pm0.04$ & $1.8^{+0.3}_{-0.3}$ & $<$0.7 & \nodata & $>$8.5 \\
1-V & $-0.33\pm0.04$ & $-0.54\pm0.08$ & $0.0555\pm0.15$ & $-0.2360\pm0.19$ & $<$ 3.9 & $<$1.2 & \nodata & \nodata \\
2-I & $1.73\pm0.10$  & $2.62\pm0.17$  & $0.0000\pm0.01$ & $0.0000\pm0.04$ & $1.7^{+0.3}_{-0.2}$ & $1.3^{+0.4}_{-0.5}$ & 49.8 & 6.2 \\
2-V & $-0.41\pm0.04$ & $-0.55\pm0.07$ & $-0.1125\pm0.01$ & $-0.0380\pm0.11$ & $<$2.6 & $<$1.8 & \nodata & \nodata\\
3-I & $5.97\pm0.33$  & $10.94\pm0.67$  & $0.0000\pm0.05$ & $0.0000\pm0.03$ & $1.7^{+0.1}_{-0.1}$ & $0.9^{+0.2}_{-0.6}$ & 93.9 & 37.9\\
3-V & $\pm0.10$        & $\pm0.10$        & \nodata & \nodata & \nodata & \nodata & \nodata & \nodata\\
4-I & $0.88\pm0.06$  & $1.94\pm0.15$  & $0.0000\pm0.09$ & $0.0000\pm0.06$ & $<$2.5 & $<$2.1 & \nodata  & $>$2.0 \\
4-V & $-0.12\pm0.03$ & $-0.42\pm0.14$ & $0.6120\pm0.57$ & $-0.5470\pm0.59$ & $<$6.5 & $<$2.9 & \nodata & \nodata\\
5-I & $1.81\pm0.10$  & $3.07\pm0.17$  & $0.0000\pm0.03$ & $0.0000\pm0.02$ & $1.6^{+0.1}_{-0.1}$ & $0.9^{+0.2}_{-0.3}$ & 46.8 & 11.3 \\
5-V & $-0.28\pm0.03$ & $-0.32\pm0.06$ & $0.8940\pm0.13$ & $0.1640\pm0.12$ & $<$1.4 & $<$1.8 & \nodata & \nodata\\
6-I & $18.78\pm1.00$ & $33.37\pm1.90$ & $0.0000\pm0.04$ & $0.0000\pm0.02$ & $<$1.9 & $<$1.9 & \nodata & $>$48.9 \\
6-V & $-0.48\pm0.06$ & $-0.76\pm0.12$ & $1.3305\pm0.12$ & $0.1830\pm0.19$ & $<$4.2 &$<$0.6 & \nodata & \nodata\\
7-I & $0.85\pm0.05$  & $1.79\pm0.13$  & $0.0000\pm0.08$ & $0.0000\pm0.05$ & $2.0^{+0.2}_{-0.1}$ & $1.3^{+0.3}_{-0.5}$ & 80.8 & 3.7\\
7-V & $-0.20\pm0.03$ & $-0.32\pm0.07$ & $-0.9045\pm0.26$ & $-0.0500\pm0.19$ & $<$2.8 & $<$2.3 & \nodata & \nodata \\
\enddata
\tablecomments{All parameters listed in this table were derived using the AIPS task \texttt{JMFIT}. For the flux density measurements in Columns~(2) and~(3), the total uncertainties were obtained by combining the formal errors with an additional 5\% systematic uncertainty in quadrature. The relative positions of Stokes $V$ emission listed in Columns~(4) and~(5) are measured with respect to the centroid of the Stokes~$I$ emission at each epoch, which is adopted as the reference origin. The absolute astrometric positions of Stokes $I$ emission were reported by \cite{ZJD2026MN}. Columns~(6)--(8) give the deconvolved sizes of major and minor axes, and position angle of the fitted elliptical Gaussian components. No Stokes~$V$ emission was detected above the $4\sigma$ level at epoch~3. We therefore report the corresponding Stokes~$V$ flux densities as $3\sigma$ limits in the table. Column~(9) lists the brightness temperatures for each epoch, or lower limits when the source was not fully resolved.}
\end{deluxetable*}

The physical emission regions of $I$ and $V$ are intrinsically extended and overlapping. The minor offsets between their centroids could help us to reveal a hint for the nonuniform distribution of the circular polarization degree in the sky plane. In Figure~\ref{fig:offsets_result}a, the offsets between the Stokes $I$ and $V$ centroids show a relatively random distribution. To remove the potential impact of the orbital motion across epochs, we took the position angle from the main to the secondary stars as the reference direction in each epoch and made another plot in Figure~\ref{fig:offsets_result}b. The orbital parameters and geometric configuration of the FF~UMa binary system, along with the corresponding spatial rotation diagram, are presented in Appendix~\ref{app:rotation}. In Figure~\ref{fig:offsets_result}b, the green data points in the quiescent state are relatively close to the secondary star. Excluding the flaring epoch (epoch~6), the remaining five epochs show $\Delta\phi$ values ranging from $-60^\circ$ to $+60^\circ$, with a mean $\Delta\phi$ of $12\,\fdg8 \pm 15\,\fdg0$, and the mean separation is $\Delta R = 13.44 \pm 3.08\,\mathrm{R}_\odot$.

\section{Discussion}
\label{sec:discussion}
\subsection{Evidence of sustained gyrosynchrotron emission in the quiescent state}
Gyrosynchrotron emission is the major mechanism to explain high-frequency radio emission in a few radio-loud RS CVn binaries \citep[e.g.][]{G_B2002,Golay2023}. Here, there is also strong evidence for the existence of sustained gyrosynchrotron emission in FF~UMa in the quiescent state. Our multi-epoch observations show that it had significant radio emission with a fractional circular polarization up to 25\% over a span of three years (cf. Figure~\ref{fig:flux_epoch}). Such moderate fractional circular polarization is not typical of coherent emission via the electron-cyclotron maser emission \citep[ECME, e.g.,][]{Slee2008,Toet2021,Vedantham2022}. We estimated the brightness temperature to be $>2.0\times10^{7}$~K. The derived values are listed in Column~(9) of Table~\ref{tab:jmfit_result}. In epochs~1, 4, and 6, the deconvolved angular sizes represent only upper limits; therefore, the derived brightness temperatures should be regarded as lower limits. The high brightness temperature and the moderate fractional circular polarization indicate that the emission at 4.8~GHz was dominated by gyrosynchrotron radiation from mildly relativistic electrons \citep{Dulk1985} in strong magnetic fields. 

Under this interpretation, we infer characteristic magnetic field strengths in the radio-emitting region of FF~UMa on the order of $40$--$160~\mathrm{G}$ (see Appendix~\ref{app:magnetic_field} for details of the estimation). This range is broadly consistent with magnetic field strengths reported for other RS~CVn-type systems \citep[e.g.][]{Owen1976ApJ,Spangler1977,Gibson1978,Estalella1993}. The comparable levels of circular polarization further suggest that FF~UMa, like other well-studied active binaries \citep[e.g.][]{Mutel1985, Mutel1987, Massi1988AA}, hosts large-scale, organized magnetic fields. Moreover, the sign of Stokes~$V$ traces the direction of the line-of-sight component of the magnetic field. The predominantly negative Stokes~$V$ detected in our observations therefore implies that the large-scale magnetic field in the radio-emitting region~(e.g., extending to $\sim40\mathrm{R}_\odot$ in epoch~7) maintains a systematic component directed away from the observer.

\subsection{Anti-correlation between circular polarization degree and radio luminosity}
\begin{figure*}[htbp]
    \centering
    \begin{subfigure}[t]{0.49\textwidth}
        \centering
        \includegraphics[height=0.42\textheight,trim=55 100 0 135, clip]{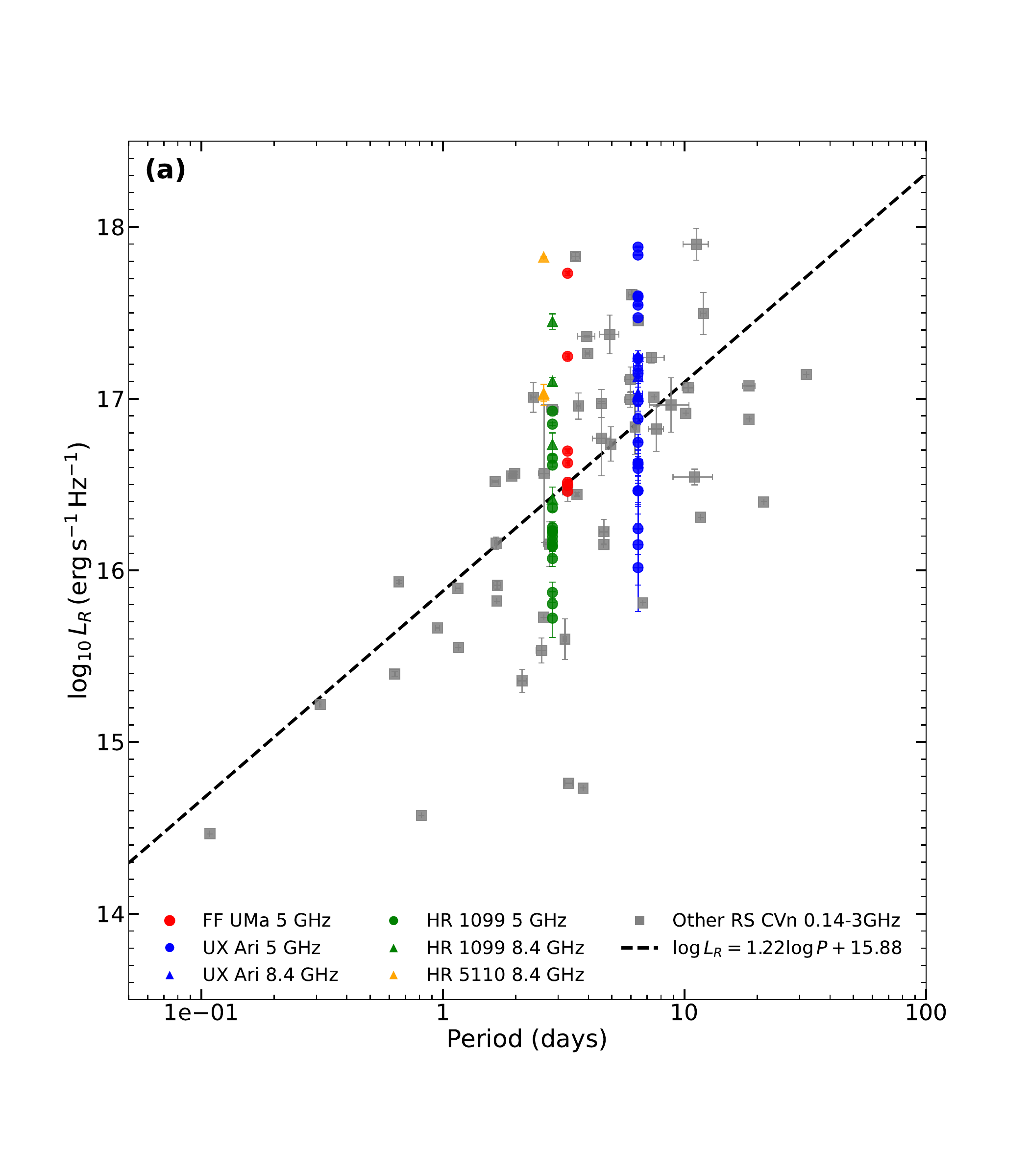}
    \end{subfigure}\hfill
    \begin{subfigure}[t]{0.49\textwidth}
        \centering
        \includegraphics[height=0.42\textheight,trim=40 50 55 135,clip]{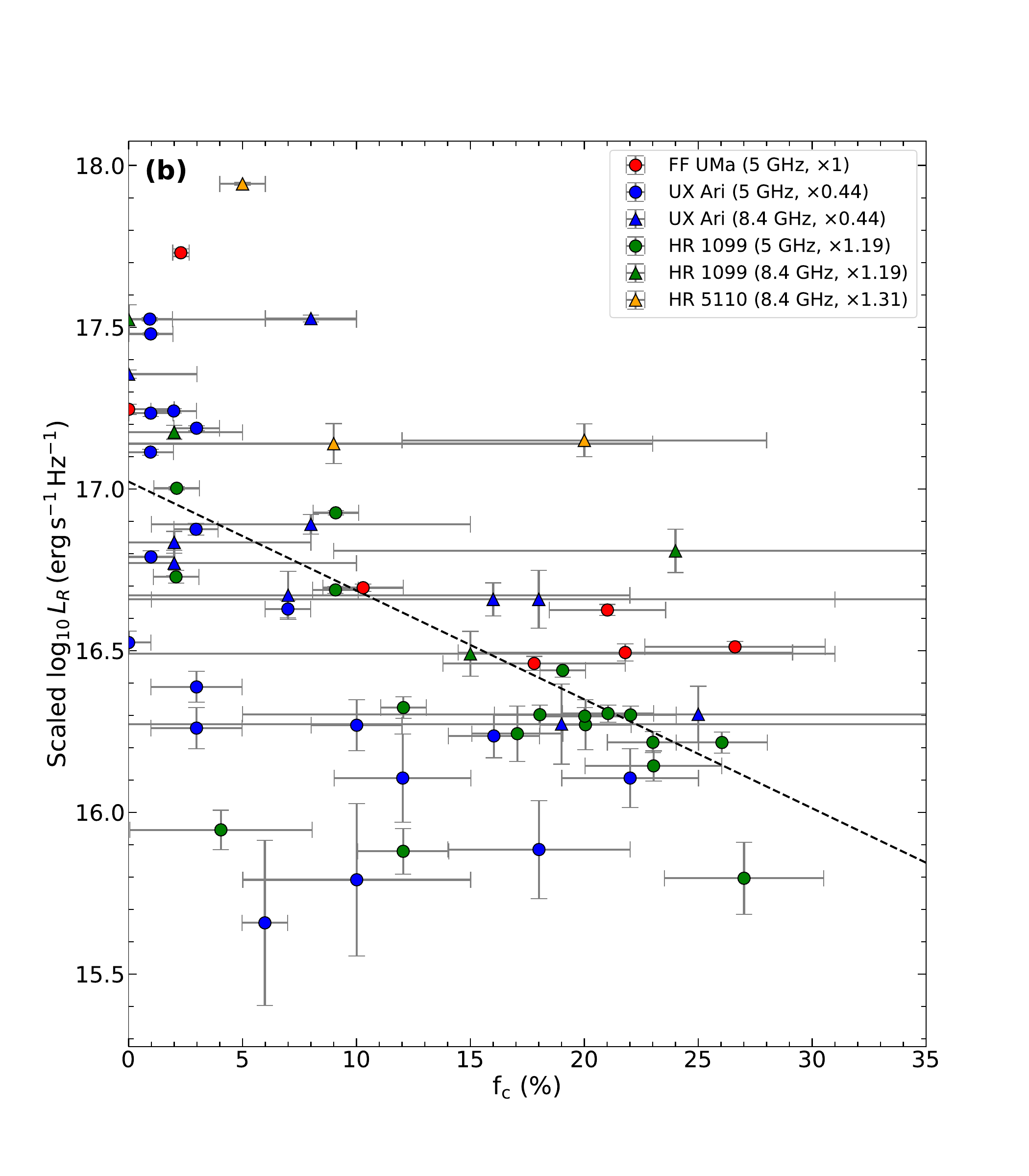}
    \end{subfigure}
    \caption{Left: Relationship between radio luminosity and orbital period for RS~CVn-type binary systems. The gray squares represent data from \citet{Huang2025}, including 42 RS~CVn-type radio stars (FF~UMa included), observed at frequencies between 144 and 3000~MHz. The shape of each symbol indicates the observing frequency: circles denote 5~GHz and triangles denote 8.4~GHz. Colors denote the sources: red for FF~UMa, green for HR~1099, blue for UX~Ari, and orange for HR~5110. The black dashed line shows the best-fitting relation obtained in this work. Right: Relationship between the scaled radio luminosity and fractional circular polarization. The symbol colors and shapes are the same as in the left panel. The scaling factor applied to the radio luminosity of each source is indicated in the legend. The black dashed line represents the best-fitting relation derived in this work.}
    \label{fig:many_stars}
\end{figure*}

Figure~\ref{fig:many_stars} shows a significant decrease in fractional circular polarization with increasing radio luminosity ($L_\mathrm{R}$) in UX~Ari, HR~1099, HR~5110 \citep{19,Mutel1987,Estalella1993}, as well as in our target FF~UMa.

According to \citet{Huang2025}, the radio luminosity of RS~CVn binaries depends on the orbital period. Using the radio flux densities reported in the sample of \citet{Huang2025} (gray points in Figure~\ref{fig:many_stars}a), together with the radio flux densities of the four binary systems collected in this work (colored points in Figure~\ref{fig:many_stars}a), we first fitted an empirical relation between the radio luminosity and the orbital period $(P)$: $L_\mathrm{R} = 10^{15.88} P^{1.22}$. Based on this empirical relation, the radio luminosities of UX~Ari, HR~1099, and HR~5110 were then rescaled to the orbital period of FF~UMa. Figure~\ref{fig:many_stars}b shows the resulting relation between the rescaled radio luminosity and the fractional circular polarization. Using uniform weighting, we fitted all the data and obtained the empirical relation:
\[
L_\mathrm{R} = 10^{-0.034\,f_\mathrm{c} + 17.02}
\]
The dashed line in Figure~\ref{fig:many_stars}b shows this fit.

The results discussed above indicate that FF~UMa is as powerful as other well-characterized RS~CVn binaries at radio wavelengths and follows a very similar anti-correlation trend. In the quiescent state, the radio emission is dominated by gyrosynchrotron radiation. The anti-correlation may indicate that, in addition to the dominant gyrosynchrotron emission, synchrotron emission produced by relativistic electrons may also contribute to the radiation. Synchrotron radiation generally does not exhibit circular polarization; therefore, when this component becomes more prominent, the total radio luminosity can increase while the overall fractional circular polarization decreases. This scenario is consistent with our observations. During the two flaring epochs (epochs~3 and 6), the radio luminosity increases significantly, while the fractional circular polarization becomes much weaker or even undetectable. Nevertheless, gyrosynchrotron emission may still operate in some regions of the magnetosphere, indicating that the observed radio emission likely arises from multiple radiation mechanisms acting simultaneously. The large-scale magnetic field might also become poorly ordered because of various flares in the corona. In addition, flaring epoch~6 may have involved an internal eruption that displaced the radio-emitting region outward, which could account for the offset position of the yellow data point in Figure~\ref{fig:offsets_result}. A similar anti-correlation was also reported in other RS~CVn systems in VLA studies \citep[e.g.][]{Garcia-Sanchez2003}.

\subsection{Additional gyrosynchrotron emission from the secondary star?}
The physical origin of the radio emission in RS~CVn binary systems remains under debate. In FF~UMa, the systematic spatial offset between the Stokes~$I$ and $V$ emission centroids in Figure~\ref{fig:offsets_result}b supports a significantly non-uniform distribution of the degree of polarization. Assuming that the Stokes~$I$ emission is dominated by the primary star, the small offsets support the possibility that there exists a certain highly polarized emission associated with the secondary star (magnetic corona and loops), the interaction regions of magnetospheres (e.g. magnetopause), or both. The centroid position of Stokes $I$ emission might also be linked to a giant magnetic loop on the surface of the primary star and thus have a large deviation ($\sim10\mathrm{R}_\odot$) from the assumed reference position in Figure~\ref{fig:offsets_result}. Because of the limited image resolutions and astrometric precisions in our observations, it is hard to directly distinguish between these possible explanations. However, there is independent support for the presence of such large-scale magnetic structures on the surface of the secondary through Doppler imaging observations \citep{Senavci2020}, which reveal a semicircular distribution of starspots at latitudes of $10^{\circ}$–$80^{\circ}$. To investigate these possible connections with the secondary, it would be necessary to carry out new VLBI observations at high frequencies in the future.       

\section{Conclusion} 
\label{sec:conclusion}
We reported multi-epoch VLBA imaging results of the RS~CVn binary system FF~UMa at 5~GHz. In addition to total intensity emission, circularly polarized emission was firmly detected in six of seven epochs, with a maximum fractional circular polarization of $26\%$. The observed high brightness temperature ($\geq10^7$ K) is also consistent with nonthermal gyrosynchrotron radiation from mildly relativistic electrons in the stellar magnetosphere. Our high-resolution studies provide strong evidence for long-lived luminous gyrosynchrotron radiation and highly-ordered magnetic fields for the first time in FF~UMa. Moreover, we noticed marginal spatial offsets between the Stokes~$I$ and $V$ emission centroids, reflecting the inhomogeneous spatial distribution of the large-scale ordered magnetic field. These results tend to support the scenario that the magnetic corona and giant loops in the secondary and magnetic interactions within the binary system also influenced the observed radio emission and its polarization properties. Thus, FF~UMa might be an ideal magnetic active binary for future high-resolution deep multi-epoch VLBI observations to explore the structure, the interaction and the evolution of large-scale magnetic fields in RS~CVn systems.  

\begin{acknowledgments}
This research is supported by the National SKA Program of China (2022SKA0120101). Y.~Gao is supported by National Natural Science Foundation of China (NSFC) under grant No. 42150105. J.~Zhang is supported by the Postdoctoral Programme for Research Institutes in Finland funded by the Finnish Government. B.~Zhang is supported by the National Key R\&D Program of China (No. 2024YFA1611501) and the Strategic Priority Research Program of the Chinese Academy of Sciences (No. XDA0350205). W.~Chen and B.~Zhang are supported by National Natural Science Foundation of China (NSFC) under grant No. 12573073, Yunnan Fundamental Research Projects (grant No. 202401AT070144) and Yunnan Foreign Talent Introduction Program (grant No. 202505AO120021). G. Gao is supported by Yunnan Fundamental Research Projects (No. 202301AT070325). Z.~Dai is supported by the Yunnan Youth Talent Project, the Yunnan Fundamental Research Projects (grant No. 202201AT070180). The National Radio Astronomy Observatory (NRAO) is a facility of the National Science Foundation (NSF), operated under cooperative agreement by Associated Universities, Inc. 
\end{acknowledgments}

\appendix
This appendix provides some additional details including the calibrator images, the geometric configuration of the binary orbit, and the procedures that we used to calculate magnetic field strength.

\section{Calibrator images}
\label{app:all_IV_fig}
\begin{figure}[ht!]
    \centering
    \includegraphics[width=0.8\linewidth, trim=0 0 90 0, clip]{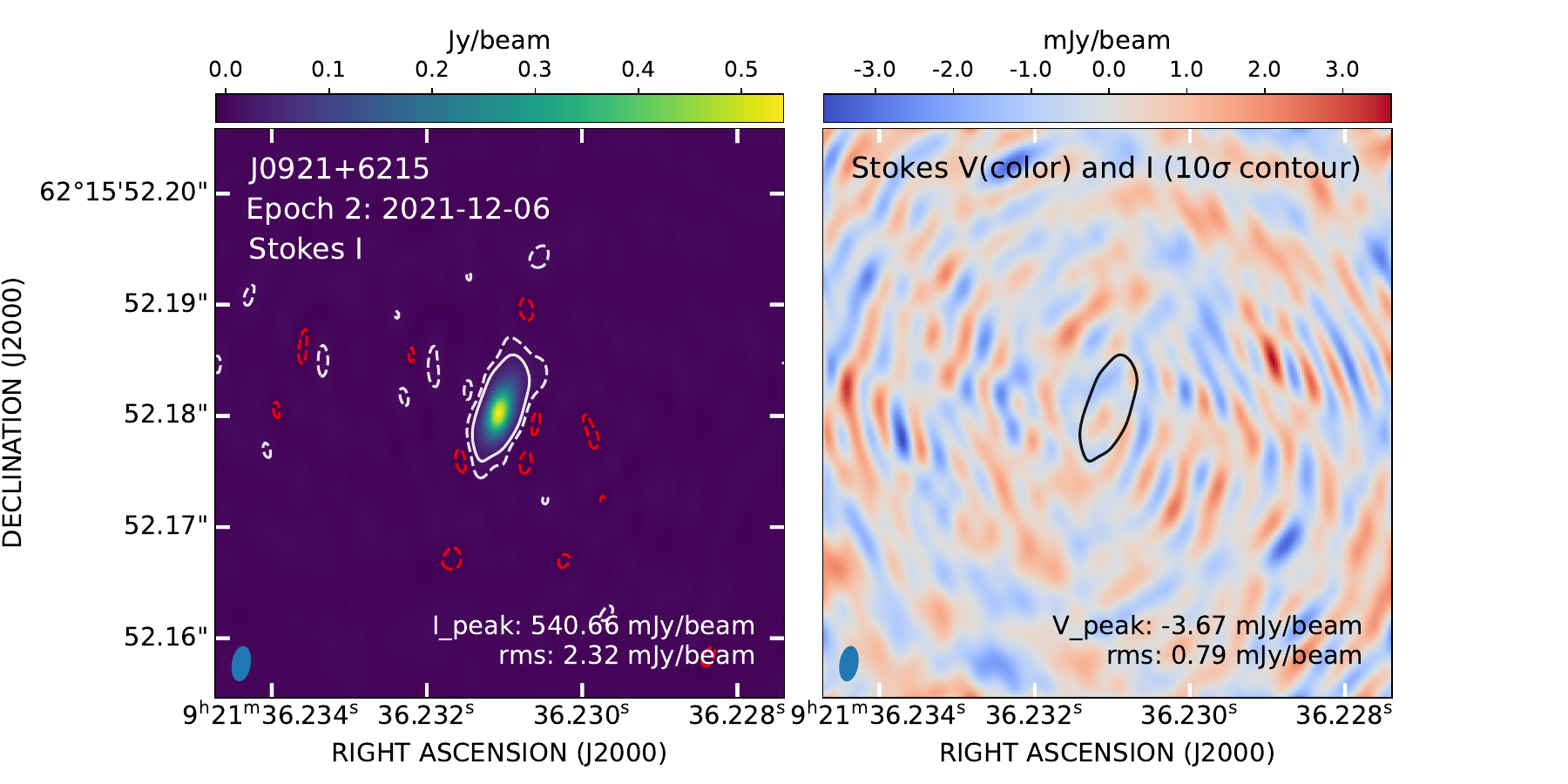}
    \caption{Calibrator J0921$+$6215 in epoch~2. Left: Stokes~$I$ image with its $+10\sigma$ (white solid), $+3\sigma$ (white dashed), and $-3\sigma$ (red dashed) contours. Right: Stokes~$V$ image overlaid with the Stokes~$I$ $+10\sigma$ (black solid) contour.
}
    \label{fig:images_of_J0921}
\end{figure}

Figure~\ref{fig:images_of_J0921} shows the images of the phase calibrator J0921$+$6215 in epoch~2. In all seven epochs, no Stokes~$V$ signal above $3\sigma$ was detected within the on-source region.

\section{The geometric relationships and spatial rotation diagram of FF~UMa}
\label{app:rotation}
\begin{figure*}[ht!]
    \centering
    \begin{minipage}{0.5\textwidth}  
        \centering
        \includegraphics[width=\linewidth, trim=0 0 100 0, clip]{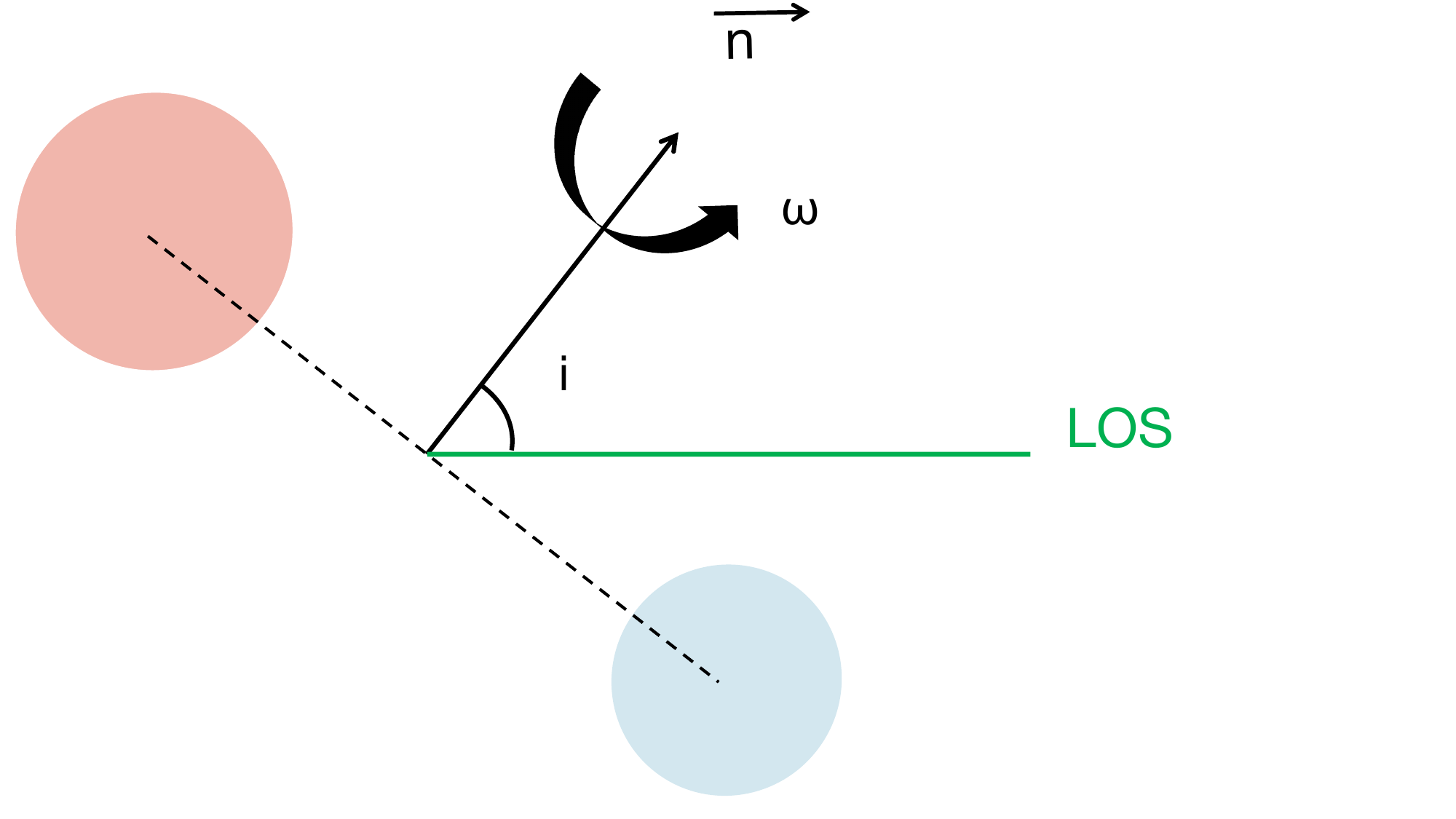}
    \end{minipage}\hfill
    \begin{minipage}{0.48\textwidth}  
        \centering
        \includegraphics[width=\linewidth]{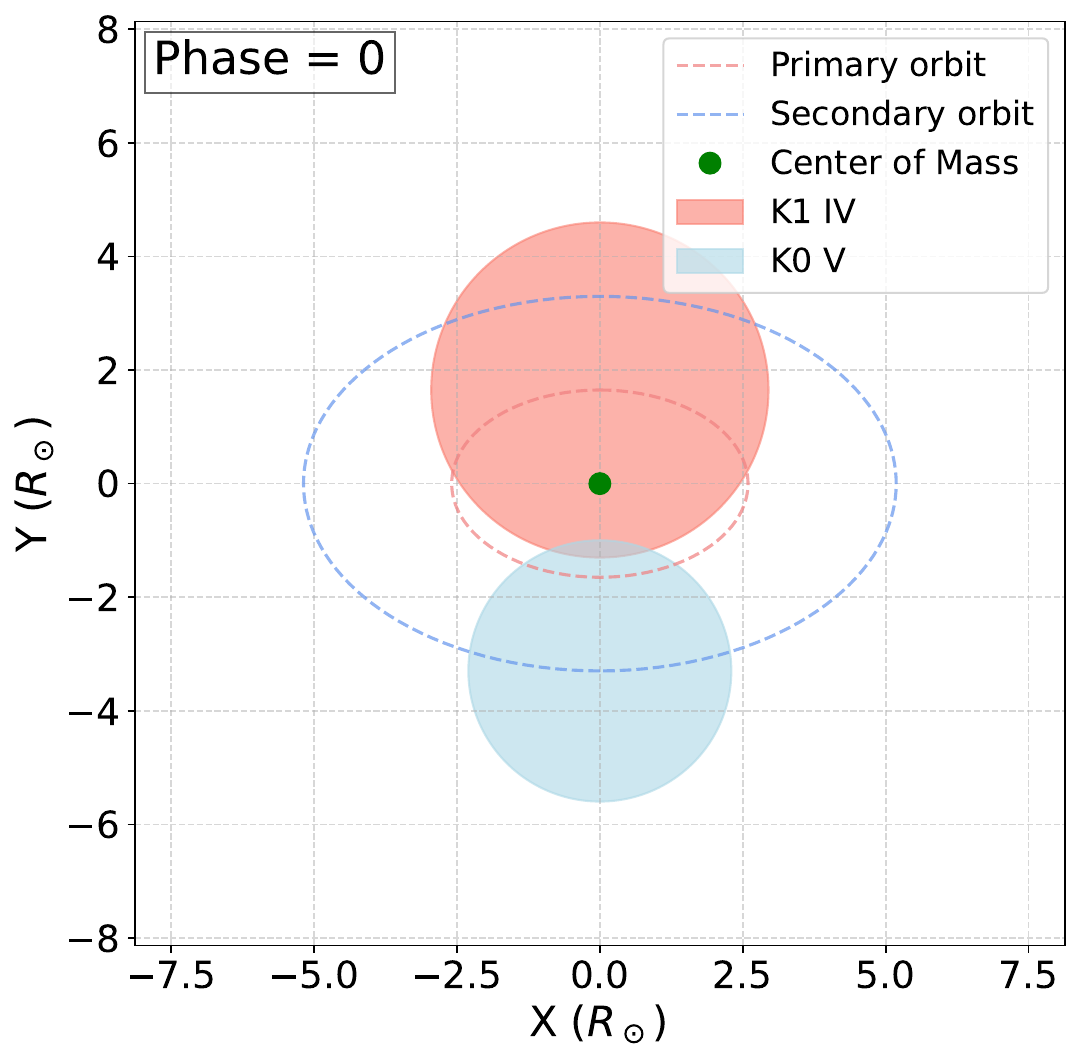}
    \end{minipage}
    \caption{Binary positions of the FF~UMa system at phase=0, with red and blue circles representing the primary and secondary stars. Left: Relative positions of the binary components. Stellar sizes and separations are schematic and do not represent true physical scales. The black dashed line indicates the orbital plane, the black straight arrow represents the orbital angular momentum vector (normal to the orbital plane), the black curved arrow shows the direction of orbital motion, and the green solid line indicates the line of sight. i denotes the orbital inclination, i=50.5$^\circ$. Right: Relative positions of the binary components projected onto the plane of the binary orbit. Physical parameters are consistent with those listed in Table~\ref{tab:orbital parameters}. Red and blue dashed ellipses represent the orbital trajectories of the primary and secondary stars, respectively.}
    \label{fig:spatial rotation}
\end{figure*}

From past optical spectroscopic studies of FF~UMa, several orbital elements of the binary system are known, such as the semimajor axis, orbital period, and eccentricity (see Table~\ref{tab:orbital parameters}). Based on \cite{Senavci2020}, and assuming the binary system rotates counterclockwise, its spatial projection on the sky plane is shown in Figure~\ref{fig:spatial rotation}. The orbital positions shown correspond to phase 0. FF~UMa is a nearly face-on binary system, and no eclipses occur along our line of sight.

\section{Magnetic field strength calculation}
\label{app:magnetic_field}
Assuming a power-law energy distribution for the emitting electrons, the effective temperature of gyrosynchrotron emission can be written as \citep[eq.~(37)][]{Dulk1985}
\begin{equation}
T_{\mathrm{eff}} \approx 2.2 \times 10^{9}
10^{-0.31\delta}
(\sin\theta)^{-0.36 - 0.06\delta}
\left( \frac{\nu}{\nu_{\mathrm{B}}} \right)^{0.50 + 0.085\delta},
\end{equation}
where $\delta$ is the electron energy spectral index, $\theta$ is the angle between the line of sight and the magnetic field direction, $\nu$ is the observing frequency in GHz, and $\nu_{\mathrm{B}}$ is the electron gyrofrequency,
\begin{equation}
\nu_{\mathrm{B}} = \frac{eB}{2\pi m_e} \approx 2.80 \times 10^{-3} B \end{equation}
with $B$ in Gauss.

For active stellar coronae, the electron energy spectral index is typically found in the range $\delta \simeq 3$--$5$ \citep[e.g.][]{Dulk1985}. In addition, the viewing angle $\theta$ between the magnetic field and the line of sight is generally unconstrained observationally; therefore, we explore a representative range of $\theta = 20^{\circ}$, $40^{\circ}$, and $60^{\circ}$.

The above approximation is valid for $10 < \nu/\nu_{\mathrm{B}} < 100$. For each combination of $\delta$ and $\theta$ within the above ranges that satisfies this condition, we estimate the corresponding magnetic field strength by equating $T_{\mathrm{eff}}$ to the observed brightness temperature. This yields characteristic magnetic field strengths in the range $B \simeq 40\text{--}160~\mathrm{G} $. 

The corresponding electron gyrofrequencies are $\nu_{\mathrm{B}} \sim 0.1$--$0.45$~GHz. We suggest that, at frequencies comparable to or below the electron gyrofrequency, coherent ECME may become important and possibly dominate, typically producing very high fractional circular polarization ($>50\%$). Indeed, \citet{Vedantham2022} reported highly circularly polarized emission from FF~UMa at 0.144~GHz with a fractional circular polarization as high as $74\%$ in right-hand circular polarization, consistent with this interpretation.

The VLBA observations analyzed here primarily probe the quiescent radio emission of FF~UMa. Significant gyrosynchrotron emission from thermal electrons would require coronal temperatures of order $\sim10^8$~K, which are typically reached only during strong radio or X-ray flares~\citep{Golay2023,Wang2026ApJ}. There is no evidence that our VLBA observations correspond to such flaring states. Under quiescent conditions, any contribution from thermal electrons is therefore expected to be negligible.

\bibliography{Ref}{}
\bibliographystyle{aasjournalv7}

\end{document}